\documentclass[prl,twocolumn,showpacs,superscriptaddress,amsmath,amssymb]{revtex4-1}
\usepackage{graphicx}
\usepackage{graphicx}

\begin{document}

\title{EuFe$_2$(As$_{1-x}$P$_x$)$_2$: reentrant spin glass and superconductivity}

\author{S.~Zapf}
\affiliation{1.~Physikalisches Institut, Universit\"at Stuttgart, Pfaffenwaldring 57, 70550 Stuttgart, Germany}
\author{H. S. Jeevan}
\affiliation{I.~Physikalisches Institut, Georg-August Universit\"at G\"ottingen, Friedrich-Hund-Platz 1, 37077 G\"ottingen, Germany}
\author{T.~Ivek}
\altaffiliation[Alt.\ address: ]{Institut za fiziku, P.O.Box 304, HR-10001 Zagreb, Croatia}
\affiliation{1.~Physikalisches Institut, Universit\"at Stuttgart, Pfaffenwaldring 57, 70550 Stuttgart, Germany}
\author{F.~Pfister}
\affiliation{1.~Physikalisches Institut, Universit\"at Stuttgart, Pfaffenwaldring 57, 70550 Stuttgart, Germany}
\author{F.~Klingert}
\affiliation{1.~Physikalisches Institut, Universit\"at Stuttgart, Pfaffenwaldring 57, 70550 Stuttgart, Germany}
\author{S.~Jiang}
\affiliation{1.~Physikalisches Institut, Universit\"at Stuttgart, Pfaffenwaldring 57, 70550 Stuttgart, Germany}
\author{D.~Wu}
\affiliation{1.~Physikalisches Institut, Universit\"at Stuttgart, Pfaffenwaldring 57, 70550 Stuttgart, Germany}
\author{P.~Gegenwart}
\affiliation{I.~Physikalisches Institut, Georg-August Universit\"at G\"ottingen, Friedrich-Hund-Platz 1, 37077 G\"ottingen, Germany}
\author{R.~K.~Kremer}
\affiliation{Max-Planck-Institut f\"ur Festk\"orperforschung, Heisenbergstrasse 1, 70569  Stuttgart, Germany}
\author{M. Dressel}
\affiliation{1.~Physikalisches Institut, Universit\"at Stuttgart, Pfaffenwaldring 57, 70550 Stuttgart, Germany}

\date{\today}

\begin{abstract}
By systematic investigations of the magnetic, transport and thermodynamic properties of single crystals of EuFe$_2$(As$_{1-x}$P$_x$)$_2$ ($0 \leq x \leq 1$) we explore the  complex
interplay of superconductivity and Eu$^{2+}$ magnetism. Below $30\,$K,
two magnetic transitions are observed for all P substituted
crystals suggesting a revision of the phase diagram. In addition to the canted A-type
antiferromagnetic order of Eu$^{2+}$ at $\sim 20\,$K, a spin glass transition is discovered at lower
temperatures. Most remarkably, the reentrant spin glass state of
EuFe$_2$(As$_{1-x}$P$_x$)$_2$ coexists with superconductivity
around $x \approx 0.2$.
\end{abstract}

\pacs{74.70.Xa,    
74.62.Fj, 
74.25.Ha, 
75.30.Gw 
75.50.Lk 
      }
\maketitle

The interplay of magnetism and superconductivity is
one of the central topics in contemporary condensed matter
research. On one hand, their antogonism is known for a century,
on the other, superconductivity was found to be closely linked to magnetism, 
for example in strongly correlated heavy fermion compounds or high-$T_c$ cuprates as well as
in the recently found iron-based pnictide or chalcogenide superconductors \cite{Johnston10,Mazin10,Norman11,Mazin11}. In the latter families, systems containing magnetic rare earth elements such as the pnictide superconductor parent compounds CeFeAsO as well as EuFe$_2$As$_2$ are of particular interest, as (besides the spin density wave in the FeAs layers) they develop an additional magnetic order of local moments at low temperatures  \cite{Zhao08,Jiang09,Xiao09}. In case of CeFeAsO, Ce$^{3+}$ antiferromagnetism appears at $\sim 4\,$K, whereas EuFe$_2$As$_2$ orders at $\sim 19\,$K with an A-type antiferromagnetic structure with the Eu$^{2+}$ moments being aligned ferromagnetically along the $a$-axis and antiferromagnetically along the $c$-axis.
EuFe$_2$As$_2$ variants are especially fascinating since despite the proximity of the  magnetic and the
superconductivity phases observed at rather high temperatures there is little variation of their transition temperatures, $T_{m}$ and $T_{c,max}$, respectively. For instance, electron doped Eu(Fe$_{1-x}$Co$_x$)$_2$As$_2$ \cite{Jiang09b}, hole doped K$_x$Eu$_{1-x}$Fe$_2$As$_2$ \cite{Jeevan08}, chemically pressurized EuFe$_2$(As$_{1-x}$P$_x$)$_2$ \cite{Jeevan11,Ren09b,Cao11} or Eu(Fe$_{1-x}$Ru$_{x}$)$_2$As$_2$ \cite{Jiao11} as well as EuFe$_2$As$_2$ under hydrostatic pressure \cite{Terashima09,Matsu11} were found to exhibit superconductivity with $T_{c,max}$ between 20 and 30 K, and simultaneously magnetic order with $T_{m}$ between 10 and 20 K.
However, up to now there is neither a clear picture how superconductivity can coexist with the strong Eu$^{2+}$ magnetism, nor a consensus  on the magnetic order in the superconducting phase.

Here we report our systematic study of the superconducting and magnetic properties
of a complete set of EuFe$_2$(As$_{1-x}$P$_x$)$_2$ ($x$ = 0, 0.055, 0.09, 0.12, 0.16, 0.165, 0.17, 0.26, 0.35, 0.39, 1) single crystals using dc and ac magnetization, dc resistivity and  heat capacity measurements. Crystals were prepared and analyzed according to standard procedures \cite{Suppl}. Magnetization data are taken in different modes: either during cooling in an applied field (FCC), or while warming up after the specimen has been cooled  in zero field (ZFC) or with a magnetic  field applied (FCH). 
For all P substituted specimen we detect two consecutive magnetic transitions separated by 1.5 K $ \lesssim \Delta T \lesssim$ 10 K requiring a revision of the phase diagram of EuFe$_2$(As$_{1-x}$P$_x$)$_2$.
Magnetic ordering at higher temperature is associated with predominant antiferromagnetic interlayer coupling, probably canted A-type antiferromagnetism, whereas the second transition at lower temperatures is identified as a spin glass-like transition evidenced by characteristic frequency dependence and thermal hysteresis effects. We conclude that the development of superconductivity is supported by the decoupling of the magnetic Eu$^{2+}$ layers in the glass phase, which could be the key to understand the interplay of superconductivity and rare eath magnetism.

\paragraph{\textbf{EuFe$_2$P$_2$}}

\begin{figure}
\centering
\includegraphics[width=.47\textwidth]{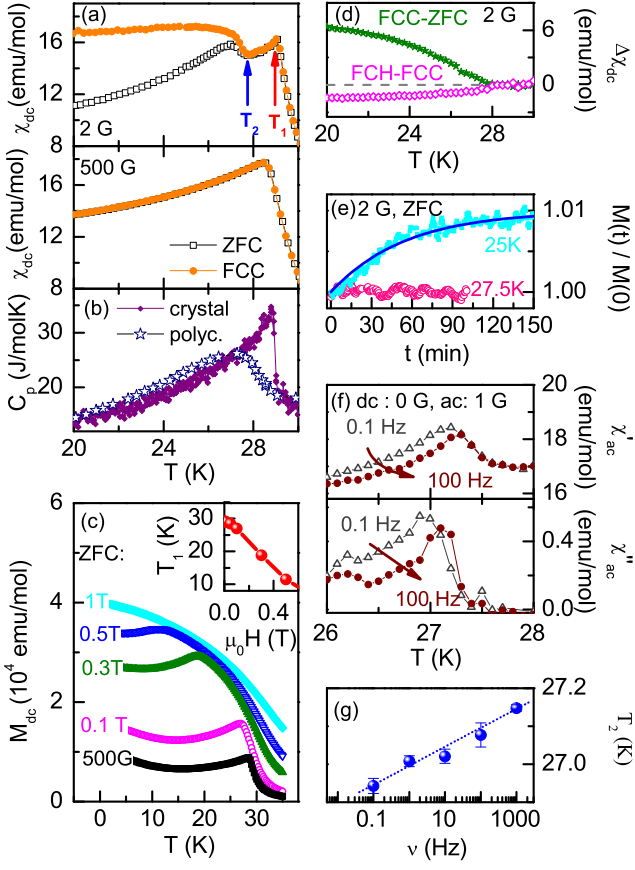}
\caption{$H || ab$ magnetization of EuFe$_2$P$_2$. (a)~ZFC (black open squares) and FCC (orange dots) curves at $2\,$G show two consecutive magnetic transitions $T_1$ and $T_2$, with a strong hysteresis at $T < T_2$ (upper panel). The transition at $T_2$ is completely suppressed with $\mu_0 H = 500\,$G (lower panel). (b)~Heat capacity for our single crystal (purple filled diamonds) and for polycrystalline EuFe$_2$P$_2$ \cite{Nowik11} (open dark blue stars) shows a broad feature covering the two magnetic transitions. (c)~ZFC magnetization for $500\,$G, $0.1\,$T, $0.3\,$T, $0.5\,$T and $1\,$T. $T_1$ decreases with increasing external field, also depicted in the inset. (d)~Magnetic hysteresis sets in at $T < T_2$, visible in $\chi_{\rm{\textit{dc,FCC}}} - \chi_{\rm{\textit{dc,ZFC}}} > 0$ (green stars) ($\mu_0 H = 2\,$G). Time dependence for $T < T_2$ is revealed by FC cycling, visible in $\chi_{\rm{\textit{dc,FCH}}} - \chi_{\rm{\textit{dc,FCC}}} < 0$ (pink open diamonds) and (e)~time-dependent magnetization after ZFC cooling ($\mu_0 H = 2\,$G, pink open circles: 27.5\,K, blue closed squares: 25K, dark blue line: fit of 25\,K data). (f)~Frequency dependence in ac susceptibility $\chi''_{ac}(T)$ and $\chi''_{ac}(T)$ (no dc field, ac drive amplitude $1\,$G, $0.1\,$Hz: dark grey open triangles, $100\,$Hz: brown dots) sets in also below $T_2$. (g)~Vogel-Fulcher-Fit of the peak below $T_2$ in $\chi''_{ac}(T)$.}
\label{Zapf_Fig1}
\end{figure}

The magnetic properties of polycrystalline EuFe$_2$P$_2$ samples have been already investigated by M\"ossbauer, specific heat and magnetization measurements \cite{Feng10}, and neutron powder diffraction \cite{Ryan11}.
Whereas Ryan \textit{et al.} interpreted their neutron diffraction data in terms of a single phase ferromagnetic transition at $\sim$30 K, Feng \textit{et al.} reported a broad smeared heat capacity anomaly. In analogy to the magnetic behavior of Eu(Fe$_{0.89}$Co$_{0.11})_2$As$_2$ \cite{Jiang09b} they tentatively analyzed their data in terms of a ferromagnetic and a subsequent ``possible helimagnetic ordering''.

Fig.~\ref{Zapf_Fig1} shows the in-plane ($H || ab$) magnetic behavior of our EuFe$_2$P$_2$ single crystal. In a very small probing field of $2\,$G two consecutive magnetic transitions can be clearly resolved in the ZFC and FCC dc magnetic susceptibility. A sharp peak at $T_{1} \approx 29\,$K is followed by an upturn starting at $T_{2} \approx 27.7\,$K leading to a second peak at $\sim 27\,$K in the ZFC magnetization [see Fig.~\ref{Zapf_Fig1}(a)].
Whereas the transition at $T_1$ exhibits no thermal hysteresis, the second transition at $T_2$ is characterized by a pronounced ZFC-FCC hysteresis which vanishes if a larger dc-magnetic field ($H || ab$) is applied. It can be finally suppressed for fields above $500\,$G. Increasing the field even higher shifts the peak at $T_1$ down [see Fig. \ref{Zapf_Fig1}(c)] and at around $1\,$T, the peak has completely disappeared. Note that a broad shoulder develops at $\sim 0.3\,$T, which for EuFe$_2$As$_2$ was interpreted  as due to a metamagnetic transition \cite{Jiang09}. Specific heat measurements on the single crystal show a sharp peak at $T_1$ with a broad shoulder at lower temperatures, consistent with the width of the anomaly for polycrystalline samples reported in Ref.~\onlinecite{Feng10} and prove that both transitions are bulk properties.
In order to get more insight into the character of the second transition at $T_2$ we studied in detail its thermal hysteresis and frequency dependence by ac magnetization measurements.
Fig.~\ref{Zapf_Fig1}(d) displays the differences between the ZFC, FCC and FCH susceptibilities for an in-plane field of $2\,$G. The thermal hysteresis is visible in the ZFC-FCC splitting at $T < T_2$. Repeated FCC and FCH cycles (heating / cooling rates 0.2 K/min) revealed a very slow time dependence of the magnetization below $T_2$ leading to a growth of the magnetization and consequently a negative difference of FCH - FCC.
A time dependence of the magnetization below $ T_2$ is also visible in a time dependence of the ZFC magnetization which after some rapid initial increase grows exponentially [see Fig.~\ref{Zapf_Fig1}(e)] with a relaxation time of $\sim$2200\,s, in good agreement with other (reentrant) spin glass systems \cite{Maji11}. The time dependence of the magnetization becomes also apparent in frequency-dependent real and imaginary component of the ac-susceptibility, $\chi'_{ac}$ and  $\chi''_{ac}$, as depicted in Fig.~\ref{Zapf_Fig1}(f). Below $T_2$ a peak appears in both components ($\chi'_{ac} \approx$ 50$\times$  $\chi''_{ac}$) which shifts to higher temperature with increasing frequency following a Vogel-Fulcher behavior [see Fig. \ref{Zapf_Fig1}(g)]. We can rule out any relation of the time and frequency dependence to flux line lattice dynamics since EuFe$_2$P$_2$ is far off from any proposed superconducting phase \cite{Jeevan11,Cao11}.

Until now magnetic ordering at higher P concentrations in polycrystalline and single crystalline samples of EuFe$_2$(As$_{1-x}$P$_x$)$_2$  was assigned to ferromagnetism \cite{Jeevan11,Cao11,Nowik11}. Spin canting with a ferromagnetic net component along the $c$-axis has been concluded from M\"ossbauer and magnetization measurements on mixed EuFe$_2$(As$_{1-x}$P$_x$)$_2$ samples \cite{Nowik11,Zapf11}.
Measurements on polycrystalline samples \cite{Cao11,Nowik11}, however, are not able to allow conclusions about possible antiferromagnetic interlayer coupling, and measurements on single crystals of EuFe$_2$(As$_{1-x}$P$_x$)$_2$ \cite{Jeevan11} failed to reveal two separate magnetic transitions and their different thermal hysteretic behavior because of too coarse temperature steps.
In view of the shape of the $M(T)$ anomaly at $T_1$ [see Fig. \ref{Zapf_Fig1}(b)] we suggest that the Eu$^{2+}$ moments in EuFe$_2$P$_2$ order rather with a canted A-type antiferromagnetic structure \cite{heli} with the spin components being ferromagnetically aligned along the $c$-axis. Additionally, below $T_2$ we detect a second phase transition with glassy character which we associate to the ordering of the in-plane components of the Eu$^{2+}$ moments.
The development of a glassy phase below a magnetic phase transition, commonly referred to as reentrant spin glass \cite{Datta84,Campbell83,Jonasen95,Enter85,Maji11,Mydosh}, indicates a  competition between antiferromagnetic and ferromagnetic spin exchange interactions in the system.
In the case of EuFe$_2$P$_2$, the antiferromagnetic RKKY interlayer coupling competes with the ferromagnetic intralayer interactions of the spins. In fact, DFT based calculations revealed a very small energy difference of antiferromagnetic and ferromagnetic ground states for EuFe$_2$(As$_{1-x}$P$_x$)$_2$ \cite{Jeevan11}. We therefore suggest that in EuFe$_2$P$_2$, competition between ferro- and antiferromagnetism causes glassy freezing of spin components in the $ab$-plane at $T_2$ and a decoupling of the magnetic Eu layers. Such a freezing of transverse magnetic compounds following long range magnetic order which has set in is consistent with mean field theoretical calculations for a reentrant spin glass \cite{Gabay81}. Our conclusions are not only supported by the time-dependent magnetization behavior at $T < T_2$, but also by the development of $M(T)$  with external fields: by application of a magnetic field of $\sim 500\,$G along the $ab$-plane, the energy barrier between different equilibrium states can be overcome, the glass transition is suppressed and the temperature dependent magnetization resembles that of EuFe$_2$As$_2$. This interpretation is consistent with the neutron powder diffraction studies by Ryan \textit{et al.} \cite{Ryan11}, as those are not sensitive to the freezing of the small in-plane spin component, as long as the ferromagnetic spin component along the $c$-axis still exists.

\paragraph{\textbf{EuFe$_2$(As$_{0.835}$P$_{0.165}$)$_2$}}

\begin{figure}
\includegraphics[width=.48\textwidth]{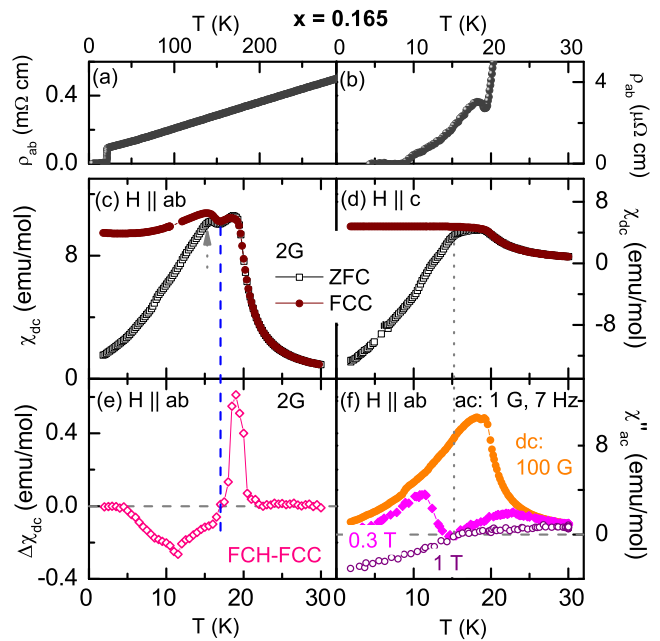}
\caption{EuFe$_2$(As$_{0.835}$P$_{0.165}$)$_2$. In-plane resistivity $\rho_{ab}(T)$ shows (a)~linear temperature dependence at high temperature and (b)~reentrant superconductivity with onset $T^{*}_{c,on} \approx 22\,$K, zero resistivity $T^{*}_{\rho=0} \approx 9\,$K. ZFC (black open squares) as well as FCC (brown dots) magnetization shows for (c)~$H || ab$ ($2\,$G) two magnetic transitions similar to $x = 1.0$ ($T_2$ is indicated by a blue dashed line), with a steep drop in the ZFC curve below $\sim 15\,$K (grey dotted arrow), which appears also for (d)~$H || c$ (grey dotted line), where negative magnetization is reached. (e)~FC cycling reveals two time-dependent glassy transitions, as FCH - FCC $>$ 0 at $T < T_{c,on}$ and FCH - FCC $<$ 0 at $T < T_{2}$. (f)~Combining ac susceptibility measurements (drive $1\,$G, frequency $7\,$Hz) with high dc fields ($100\,$G: orange dots, $0.3\,$T: pink diamonds, $1\,$T: purple open circles), negative magnetization is revealed for $H || ab$ below $15\,$K (grey dotted line).}
\label{Zapf_Fig2}
\end{figure}

In order to study the complex interplay of magnetism and superconductivity in mixed As - P samples we have investigated in detail the magnetic and superconducting properties of a single crystal of EuFe$_2$(As$_{0.835}$P$_{0.165}$)$_2$. The in-plane electrical resistivity [see Fig.~\ref{Zapf_Fig2}(a)] proves onset of superconductivity at $T^{*}_{c,on} \approx 22\,$K indicated by  a steep initial decrease of the resistivity. Zooming into the transition reveals reentrant behavior at about $19\,$K followed by a smooth decrease towards zero resistivity  which is achieved only below $T^{*}_{\rho=0} \approx 9\,$K [see Fig.~\ref{Zapf_Fig2}(b)].
Fig.~\ref{Zapf_Fig2}(c)-(f) compile selected ac and dc magnetization data obtained with experimental configurations identical to those used for EuFe$_2$P$_2$. The dc magnetizations at low fields, similar to those of EuFe$_2$P$_2$, show two peaks for $H || ab$ which are shifted to lower temperatures [$T_1 \approx 19\,$K, $T_2 \approx 16.8\,$K, see Fig.~\ref{Zapf_Fig2}(c)]. A steep downturn occurs below 15\,K for both, $H || c$ and $H || ab$ magnetizations, which ends up in a diamagnetic signal for $H || c$, indicating superconducting shielding [see Fig.~\ref{Zapf_Fig2}(d)].  As $H || ab$ requires shielding currents perpendicular to the layers, the magnetization stays positive. However, performing in-plane ac susceptibility measurements with applied high dc fields $H || ab$ [see Fig.~\ref{Zapf_Fig2}(e)]  reveals also a diamagnetic in-plane shielding signal if the dc field is large enough ($\gtrsim$ 1 T) to saturate the Eu$^{2+}$  magnetism.

Repeated field cooled (FC) cycling again reveals a time dependence of the magnetization with a negative FCH - FCC difference below $T < T_{2}$ characteristic of  glassy magnetism, similar to that found in EuFe$_2$P$_2$ [see Fig.~\ref{Zapf_Fig1}(c)]. We therefore conclude that superconducting EuFe$_2$(As$_{0.835}$P$_{0.165}$)$_2$  shows an analogous reentrant spin glass behavior as EuFe$_2$P$_2$. The additional positive peak in the FCH-FCC curve between 17 and 21\,K could be ascribed to vortex dynamics, as it coincides with the steep initial decrease of the resistivity marking the onset of superconductivity.

\paragraph{\textbf{Phase Diagram}}
\begin{figure}
\includegraphics[width=.48\textwidth]{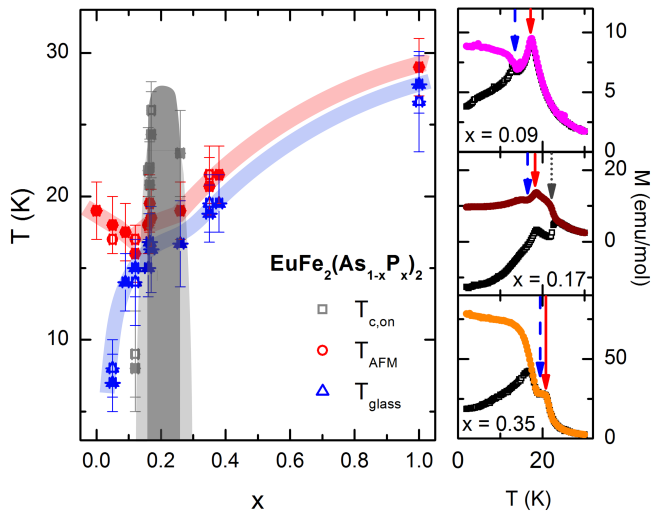}
\caption{Phase Diagram of EuFe$_2$(As$_{1-x}$P$_x$)$_2$. $T_{cAFM}$ (red dots) indicates a canted A-type antiferromagnetic transition, with a ferromagnetic net component of the Eu$^{2+}$ spins along the $c$-direction, $T_{glass}$ (blue triangles) a spin glass transition due to the freezing of the spins in the $ab$-plane and $T_{c,on}$ the onset of superconductivity (grey squares). Closed symbols indicate transition temperatures deduced from magnetization, open ones from resistivity measurements. Shadowed lines are guides to the eyes. The light grey area indicates the onset of superconductivity, while bulk superconductivity is fully developed in the dark grey regime \cite{Jeevan11,Tokiwa12}. The right panel shows typical corresponding $M(T)$ curves. Non-superconducting $x = 0.09$ (ZFC: open squares, FCC: pink dots), superconducting $x = 0.17$ (ZFC: open squares, FCC: brown dots) as well as non-superconducting $x = 0.35$ (ZFC: open squares, FCC: orange dots) samples are selected. Arrows indicate the transition temperatures for the antiferromagnetic (red solid lines), spin glass (blue dashed lines) and superconducting (grey dotted line) phases.}
\label{Zapf_Fig3}
\end{figure}

In order to follow compositional  dependence of the two magnetic transitions consistently found in EuFe$_2$P$_2$ and EuFe$_2$(As$_{0.835}$P$_{0.165}$)$_2$, we have extended our studies to EuFe$_2$(As$_{1-x}$P$_x$)$_2$ single crystals with $x$ = 0, 0.055, 0.09, 0.12, 0.16, 0.17, 0.26, 0.35 and 0.39 (see Fig.~\ref{Zapf_Fig3}). In all P substituted specimen we observe two consecutive magnetic transitions, which we ascribe, in analogy to the previous sections, to a canted A-type antiferromagnetic transition with the Eu$^{2+}$  spin components along the $c$-axis being ferromagnetically aligned below $T_1 = T_{cAFM}$ and to a glassy freezing of the spin components in the $ab$-plane at $T_2 = T_{glass}$, with $T_{cAFM} > T_{glass}$.  In Fig.~\ref{Zapf_Fig3} we have compiled  the resulting magnetic phase diagram together with the  superconducting dome.

According to our investigations on single crystals the reentrant spin glass transition appears for all P substituted specimen. The effect of chemical disorder of the As and P anions on the RKKY exchange must be ruled out as the origin of the glass transition since it occurs also in well ordered EuFe$_2$P$_2$ crystals. We rather ascribe the glass transition to competition of ferromagnetic interactions within a layer with antiferromagnetic RKKY interactions between neighboring layers.

The transition temperatures exhibit a nonmonotonic behavior with P substitution. At low P concentration, $0 <  x \lesssim 0.12$, the antiferromagnetic Eu$^{2+}$ transition temperature follows the transition temperature of the spin density wave. Coupling of the itinerant Fe magnetism and the Eu$^{2+}$ local spin moments was theoretically predicted and experimentally confirmed by the increasing canting of spins out of the $ab$-plane concomitant with the suppression of the spin density wave. \cite{Akbari11,Akbari13,Nowik11,Zapf11} With  increasing canting, the ferromagnetic component of the Eu$^{2+}$ along the $c$-direction increases,  and  the competition with the antiferromagnetic RKKY interaction between the layers is enhanced which finally leads to the development of the spin glass phase. In the superconducting regime, the transition temperatures vary only slightly with P concentration. When superconductivity is finally suppressed, both transition temperatures, $T_1$ and $T_2$ increase markedly, probably due to a Lifshitz transition \cite{Maiwald12,Thirupathaiah11,Tokiwa12} which effects the RKKY exchange.

Antiferromagnetic interlayer coupling developing up to high P concentrations, as well as a rather narrow superconducting dome, are consistent with experiments on EuFe$_2$As$_2$ under pressure \cite{Matsu11}. Between the concentrations $x \approx$ 0.12 and $x \approx$ 0.26 the onset of a superconducting transition is found, while fully developed bulk superconductivity occurs in an even narrower regime \cite{Jeevan11,Tokiwa12}. As concluded previously, a Lifshitz transition near $x \approx$ 0.23 coincides with the upper limit of superconductivity \cite{Maiwald12,Thirupathaiah11,Tokiwa12}. Investigations of polycrystalline samples, however, resulted in a somewhat broader dome extending to an upper limit of $x \approx$ 0.4 \cite{Nowik11,Cao11}. The assignment of the upper limit was based on the assumption that two subsequent resistivity anomalies seen in samples with $x \approx$ 0.4 indicate onset of superconductivity succeeded by reentrance due to ordering of the Eu$^{2+}$ spins. Our experiments on single crystals rather indicate that these two anomalies are purely of magnetic origin as we do not see any signature of superconductivity in our $x$=0.35 crystal (see Fig.~\ref{Zapf_Fig3}, right panel) \cite{PME2}.

The question how bulk superconductivity can coexist with Eu$^{2+}$ magnetic ordering is quite fundamental and requires the exact knowledge of the magnetic structure.
Consistent with M\"ossbauer and neutron powder diffraction \cite{Nowik11,Ryan11}, the results of our experiments imply that a large net component of the Eu$^{2+}$ spins is ferromagnetically aligned perpendicular to the layers. In addition, we find that glass-like dynamics and freezing of the in-plane component develops below $T_2$ which destroys coherence between the Eu layers. Superconductivity in the iron based superconductors is commonly believed to take place mainly in the FeAs layers. In this scenario, the inner field resulting from the Eu$^{2+}$ ferromagnetic component along the $c$-axis could be screened by the formation of spontaneous vortices perpendicular to the layers \cite{Jiao11}. Together with the destroyed coherence between the Eu layers due to the glass dynamics this scenario might be key to how superconductivity can coexist with the usually strong Eu$^{2+}$ magnetism.
It will be therefore very interesting to investigate also other Eu containing pnictides in more detail in order to understand whether the glass phase is required for the existence of superconductivity in these systems.

\paragraph{\textbf{Conclusion}}
EuFe$_2$(As$_{1-x}$P$_x$)$_2$ exhibits two consecutive magnetic transitions $T_{cAFM} > T_{glass}$ over the entire P substitution range.  From magnetization data, we identify the higher temperature transition as a canted A-type antiferromagnetic transition. The spin canting increases with P substitution concomitant with the suppression of the spin density wave, until the spins are aligned almost along the $c$-direction, $i.e.$ ferromagnetic intralayer coupling competes with antiferromagnetic RKKY interlayer coupling. This causes glassy behavior of the spin components in the $ab$-plane at $T_{glass}$, evidenced by characteristic frequency- and time-dependent response of the magnetization. Thus EuFe$_2$(As$_{1-x}$P$_x$)$_2$ is a reentrant spin glass and does not sustain conventional antiferromagnetic or ferromagnetic Eu$^{2+}$ magnetic ordering down to low temperatures. Development of superconductivity is supported by the decoupling of the magnetic Eu$^{2+}$layers.

We thank R. Beyer, J. R. O' Brien, I. Eremin, and I. \v{Z}ivkovi\'{c} for fruitful discussions and E. Br\"ucher, G. Siegle, and G. Untereiner for expert experimental assistance. This work was supported by DFG SPP 1458.


\begin{thebibliography}{99}
\bibitem{heli}
Due to sample shape effects we can not distinguish between A-type antiferromagnetism and helimagnetism.
\bibitem{PME2}
The double-feature in $M(T)$ curves for superconducting samples was in some cases assigned to result from the Paramagnetic Meissner Effect \cite{Nowik11}. However, this effect should only appear in the FC curve \cite{PME}, whereas features are observed also in ZFC data.
\end{thebibliography}
\end{document}


\textbf{Supplemental Material}
\\

\paragraph{\textbf{Experimental techniques}}
EuFe$_2$(As$_{1-x}$P$_x$)$_2$ high quality single crystals were synthesized by the Bridgeman method \cite{Jeevan11} and studied using a Quantum Design $7\,$T MPMPS-XL superconducting quantum interference device (SQUID). Magnetization data are complemented with four point dc resistivity measurements with a current of the typical value $1\,$mA. In the case of small resistances (especially for the $c$-direction measurements), lock-in technique with low frequencies ($77\,$Hz) was used. Heat capacities were measured with a relaxation-type calorimeter.
\\

\paragraph{\textbf{Magnetization measurements}}
In order to reveal glass-like behavior, special care has to be taken concerning the cooling procedure. When measuring at stabilized temperatures $T$, the stabilization process usually also includes oscillating around the target temperature and its duration dependends on the setpoint temperature. For a material with time-dependent magnetization $M$, both lead to non-reproducibile $M(T)$ curves. Therefore we typically chose to sweep the temperature in $M(T)$ measurements. For the dc magnetization we used the sweep velocity 0.2\,K/min.  Typical measurements for EuFe$_2$(As$_{1-x}$P$_x$)$_2$ take less than 10\,s with the settings: 4\,cm scan length, 2\,scans per measurements and 24\,points, $i. e.$ the temperature error is less than 0.05\,K and thus negligible (visible also in the absent thermal hysteresis in the paramagnetic regime). For ac measurements covering a big temperature range, we chose the sweep velocity 0.1\,K/min, as here measurements usually are slower. With the parameters: 1\,G drive amplitude, 7\,Hz wave frequency, amplifier gain 1, 1\,block to average, 1\,scan per measurement and 1\,s settling time, measurements were faster than 1\,min, $i. e.$ in this case the temperature error is less than 0.1\,K. For frequency dependent ac measurements around the spin glass transition, however, we had to settle the temperature and measure for one temperature different frequencies. Frequency dependent susceptibility is found no matter whether frequencies are raised or decreased, meaning that the slow increase of the spin glass magnetization with time has a smaller effect than the frequency dependence of the susceptibility.

AC measurements were always done in the ZFC (zero-field-cooling) mode. In this mode, the sample is cooled down in zero net magnetic field, at lowest temperatures (in our case 2\,K) the magnetic field is eventually switched on (with the ``no overshoot mode'') and then the measurement is performed while heating. For dc measurements, additional FC (field-cooled) measurements were performed with the sample being cooled in the external field at which the measurement is done. Here one has to distinguish between FCC (field-cooled-cooling), $i. e.$ the magnetization is measured while cooling, and FCH (field-cooled-heating), $i.e.$ the sample is cooled with an applied external magnetic field, but the magnetization is measured then while heating up. Thus the usual procedure for our dc measurements includes ZFC, FCC and FCH curves, whereas in literature often only ZFC and FCC curves are measured. However, FC cycling ($i.e.$ measuring FCH and FCC curves) can reveal important information if the sample exhibits time dependent magnetization, as discussed in our manuscript.

As the dc magnetic field is produced by superconducting coils, one has to consider the effect of the remanent magnetization captured in them. Usually, we therefore determine the zero field by searching at temperatures above the superconducting and Eu$^{2+}$ magnetic transitions the magnetic field (in the range of a few G), which leads to zero signal at the SQUID. However, as the cooling process in a homogeneous zero field can be crucial especially for superconductors \cite{QDUSA}, for $x = 0.165$ and $0.35$ $\mu_0 H = 2\,$G measurements were also done using the Quantum Design``ultra low field'' option which guarantees field nulling better than $0.05\,$G remanent field. We did not find any qualitative difference of the magnetization compared to measurements with our standard procedure to determine the zero field.
\\

\paragraph{\textbf{Investigated samples}}
Table \ref{mass} summarizes the compositions $x$ and masses $m$ of EuFe$_2$(As$_{1-x}$P$_x$)$_2$ single crystals that were investigated.

\begin{table}[!h]
  \centering
  
  \begin{tabular}{ | p{1cm} ||  p{1cm} | p{1cm} | p{1cm} | p{1cm} | p{1cm} | p{1cm} | }
    \hline
    \textit{\textbf{x}} & 0 & 0.055 & 0.09 & 0.12 & 0.16 & 0.165 \\ \hline 
    \textit{\textbf{m}}(mg) & 6.1 & 9.7 & 0.45 & 1.1 & 2.3 & 7.0 \\ \hline 
  \end{tabular}

  \begin{tabular}{ | p{1cm} ||  p{1cm} | p{1cm} | p{1cm} | p{1cm} | p{1cm} | p{1cm} } 
  \cline{1-6} 
    \textit{\textbf{x}} & 0.17 & 0.26 & 0.35 & 0.39 & 1 & \\ \cline{1-6} 
   \textit{\textbf{m}}(mg) & 1.2 & 0.44 & 0.59 & 4.0 & 0.16 & \\ \cline{1-6}
  \end{tabular}

  \caption{Compositions $x$ and masses $m$ of investigated EuFe$_2$(As$_{1-x}$P$_x$)$_2$ single crystals.}
  \label{mass}
\end{table}